\RequirePackage{fix-cm}
\documentclass[twocolumn]{svjour3}          % twocolumn
\smartqed  % flush right qed marks, e.g. at end of proof
\usepackage{graphicx}
\graphicspath{{./figs/}}
\usepackage{mathptmx}      % use Times fonts if available on your TeX system
\usepackage[round]{natbib}
\usepackage[version=3]{mhchem}
\usepackage{color}
\usepackage{amsmath}
\usepackage{booktabs,caption}
\usepackage[flushleft]{threeparttable}
\usepackage{multirow} 
\usepackage{fixltx2e}
\usepackage{nicefrac}

\newcommand{\AmTh}{NH\textsubscript{4}\-SCN}
\newcommand{\AmThSol}{NH\textsubscript{4}\-SCN so\-lu\-tion}

\journalname{Experiments in Fluids}
\begin{document}

\title{Aqueous ammonium thiocyanate solutions as refractive\\
index-matching fluids with low density and viscosity%\thanks{Grants or other notes
}


\author{D. Borrero-Echeverry\textsuperscript{1,\,2}        \and
	B.C.A. Morrison\textsuperscript{2} %etc.
}

\authorrunning{D. Borrero-Echeverry and B.C.A. Morrison} % if too long for running head

\institute{ D. Borrero-Echeverry\\
							\email{dborrero@willamette.edu}
							\vspace{6pt}
							\at
              \textsuperscript{1}\,Department of Physics, Willamette University, 900 State Street, Salem, OR 97301, USA
						  \and 
								\at \textsuperscript{2}\,Department of Physics, Reed College, 3203 SE Woodstock Boulevard, Portland, OR 97202, USA
 %            \emph{Present address:} of F. Author  %  if needed
}

\date{Received: date / Accepted: date}

\maketitle

\begin{abstract}
We show that aqueous solutions of ammonium thiocyanate (\AmTh) can be used to match the index of refraction of several transparent materials commonly used in experiments, while maintaining low viscosity and density compared to other common refractive index-matching liquids. We present empirical models for estimating the index of refraction, density, and kinematic viscosity of these solutions as a function of temperature and concentration. Finally, we summarize the chemical compatibility of ammonium thiocyanate with materials commonly used in apparatus. 

% 78.15.+e - Optical properties of fluid materials, supercritical fluids and liquid crystals
% 83.85.Jn - Viscosity measurements
% 47.80.Jk - Flow visualization and imaging
\end{abstract}

\section{Introduction}
\label{intro}
In recent decades, particle image velocimetry (PIV), laser Doppler anemometry (LDA), and laser-induced fluorescence (LIF) have become standard tools in the experimentalist's toolbox. A common problem that arises when using these techniques is that as light passes through the various interfaces of the experimental apparatus, it refracts, leading to distorted images that are difficult to analyze. For measurements in liquids, a common solution to this problem is to use refractive index-matching fluids to minimize these effects \citep{Budwig1994}.

Typical refractive index-matching fluids include mixtures of halogenated hydrocarbons with organic solvents or aqueous solutions of heavy ionic salts \citep{Donnelly1981}. Most of these fluids have viscosities greater than that of water making it difficult to achieve high Reynolds numbers in experiments. They can also have relatively high specific gravities, which makes it challenging to find density-matched tracers for particle-based measurements. As we will show, ammonium thiocyanate (\AmTh) solutions have indices of refraction that match those of transparent materials frequently used in the construction of apparatus, while having physical properties much closer to those of water. This makes them useful when employing optical techniques to study high Reynolds number flows in complicated geometries.

Aqueous ammonium thiocyanate solutions have previously been used as refractive index-matching media by several authors, including Budwig and his collaborators, who used them to aid flow visualization within aortic aneurysm models \citep{Budwig1993,Egelhoff1999}. More recently, \AmThSol s have been used in PIV studies of flow through randomly packed porous beds \citep{Patil2013} and in tomographic PIV studies of turbulent structures in Taylor-Cou\-ette flow \citep{Borrero2014Thesis}. However, data on the optical, physical, and chemical properties of \AmThSol s are not well-do\-cu\-men\-ted in the literature. We aim to fill this gap and provide a useful guide for researchers to develop \AmTh-based refractive index-matching fluids. 

Our discussion will be limited to pure solutions of \AmTh, which result in index-matching fluids with the low viscosities and densities that are of interest to us. However, \AmThSol s can also be used as a starting point in formulating refractive index-matching fluids with other design criteria. A good example of this is provided by \citet{Bailey2003}, who document a procedure for developing refractive index- and density-matched solutions for use with PMMA particles using ternary mixtures of water, \AmTh, and glycerin. Their approach mimics earlier work \citep{Jan1989,Budwig1997} where glycerin was used to tune the viscosity of the working fluid, while maintaining its useful refractive index-matching properties. 

The remainder of this Letter is organized as follows: Section \ref{sec:Optical} discusses the optical properties of \AmThSol s. Section \ref{sec:Physical} discusses the dependence of their viscosities and densities on temperature and \AmTh\, concentration. Section \ref{sec:Chemical} discusses the chemical compatibility of \AmTh\ with materials commonly used to build experimental apparatus, as well as toxicity and handling information. Finally, Section \ref{sec:Summary} provides a summary of our results.

\section{Optical Properties of \AmThSol s}
\label{sec:Optical}

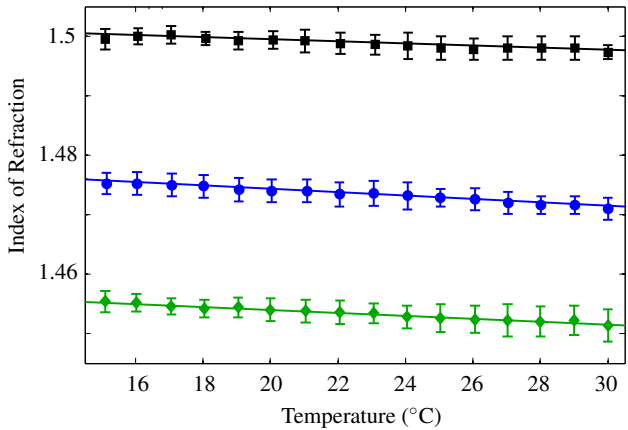
\begin{figure}
  \centering
  \setlength{\unitlength}{0.44\textwidth}%
\begin{picture}(1,0.66131547)%
    \put(0,0){\includegraphics[width=\unitlength]{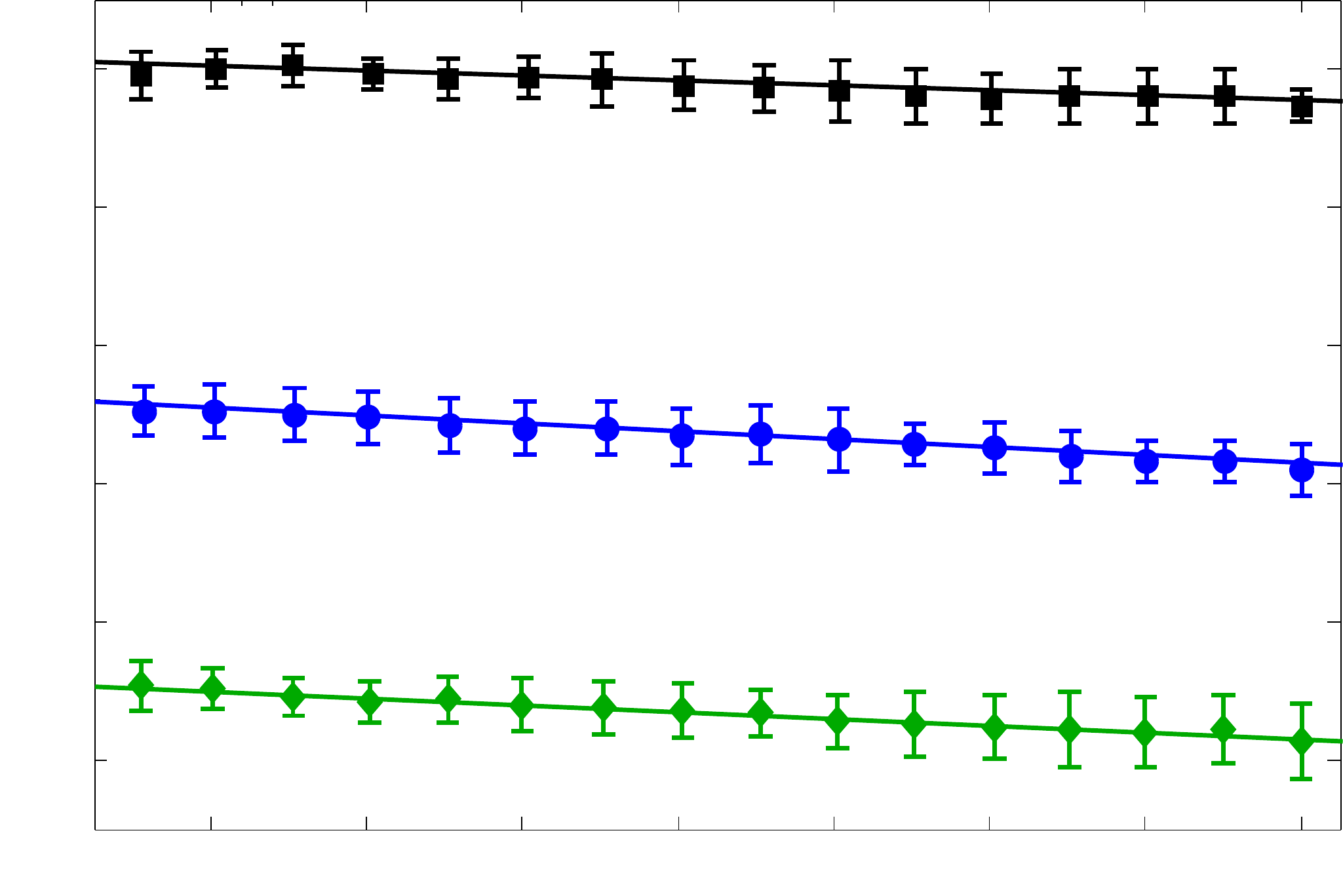}}%
    \put(0.14129077,0.00093051){\makebox(0,0)[lb]{\smash{16}}}%
    \put(0.25598354,0.00093051){\makebox(0,0)[lb]{\smash{18}}}%
    \put(0.37067866,0.00093051){\makebox(0,0)[lb]{\smash{20}}}%
    \put(0.48663247,0.00093051){\makebox(0,0)[lb]{\smash{22}}}%
    \put(0.60132591,0.00093051){\makebox(0,0)[lb]{\smash{24}}}%
    \put(0.71601935,0.00093051){\makebox(0,0)[lb]{\smash{26}}}%
    \put(0.83071279,0.00093051){\makebox(0,0)[lb]{\smash{28}}}%
    \put(0.94666827,0.00093051){\makebox(0,0)[lb]{\smash{30}}}%
    \put(-0.010455519,0.19581975){\makebox(0,0)[lb]{\smash{1.46}}}%
    \put(-0.010455519,0.40000096){\makebox(0,0)[lb]{\smash{1.48}}}%
    \put(0.0058985,0.60418049){\makebox(0,0)[lb]{\smash{1.5}}}%
    \put(0.40804034,-0.05424389){\makebox(0,0)[lb]{\smash{Temperature ($^\circ$C)}}}%
    \put(-0.03468131,0.23888088){\color[rgb]{0,0,0}\rotatebox{90}{\makebox(0,0)[lb]{\smash{Index of Refraction}}}}%
  \end{picture}%
	\vspace{10pt}
  \caption{\label{fig:NVsT} The index of refraction of \AmThSol s decreases approximately linearly with temperature for fixed concentrations of 49.8\% (green diamonds), 55.1\% (blue circles), and 62.6\% (black squares) \AmTh\ by weight. Error bars indicate the sample standard deviation of the index of refraction at a given temperature and concentration.}
\end{figure}

\begin{table}
\centering
  \begin{threeparttable}
    \caption{\label{tab:RefractionFitsT}Fit parameters for index of refraction versus temperature $T$ {models at} fixed concentration, $n(T) = n_0 + n_1\,T$. Here, $r^2$ is the coefficient of determination.}
     \begin{tabular}{ccccc}
      \toprule
       Concentration & $n_0$ & $n_1$ & $r^2$  \\
       (\% by weight)&  & ($^\circ$C$^{-1}$) & \\
      \midrule
					62.6 & 1.503 & $-1.764 \times 10^{-4}$ & 0.910 \\
					55.1 & 1.480 & $-2.832 \times 10^{-4}$ & 0.979 \\
					49.8 & 1.459 & $-2.506 \times 10^{-4}$ & 0.976 \\      
			\bottomrule
    \end{tabular}
  \end{threeparttable}
\end{table}

We began our experiments by preparing solutions with concentrations of 49.8\%, 55.1\%, and 62.6\% \AmTh\ by weight as determined using an analytical scale. These concentrations were chosen because preliminary experiments showed that at room temperature their refractive indices match those of fused quartz, borosilicate glass, and acrylic, respectively. The solutions were prepared with 99+\% pure ammonium thiocyanate (available for $\sim$ US\$100 per kg from Acros Organics) and de-ionized water. Because the solvation of \AmTh\ is endothermic, the solution was gently heated as \AmTh\ was added to help it dissolve more quickly. Despite the purity of the \AmTh\ used, some insoluble impurities remained, which made the solutions slightly cloudy. These impurities were removed using filter paper. 

The index of refraction $n$ of \AmThSol s at 589.3 nm was measured as a function of temperature using a Bausch \& Lomb Abbe-3L refractometer. A recirculating heat bath allowed us to adjust the temperature of the refractometer and solution to within 0.01$^\circ$C. Measurements of the index of refraction of each solution were taken at 1$^\circ$C intervals from 15$^\circ$C to 30$^\circ$C with five runs taken at each concentration. It was determined that, as with sodium iodide-based index-matching fluids \citep{Narrow2000}, the index of refraction of \AmThSol s decreases approximately linearly with increasing temperature (at fixed concentration) for the range parameters studied. The data at each temperature and concentration were averaged and the aggregate data was fit to linear regression models, as shown in Fig.~\ref{fig:NVsT}. The fit parameters for the three data sets are summarized in Table \ref{tab:RefractionFitsT}.

\begin{figure}
  \centering
  \setlength{\unitlength}{0.44\textwidth}%
    \begin{picture}(1,0.66719507)%
    \put(0,0){\includegraphics[width=\unitlength]{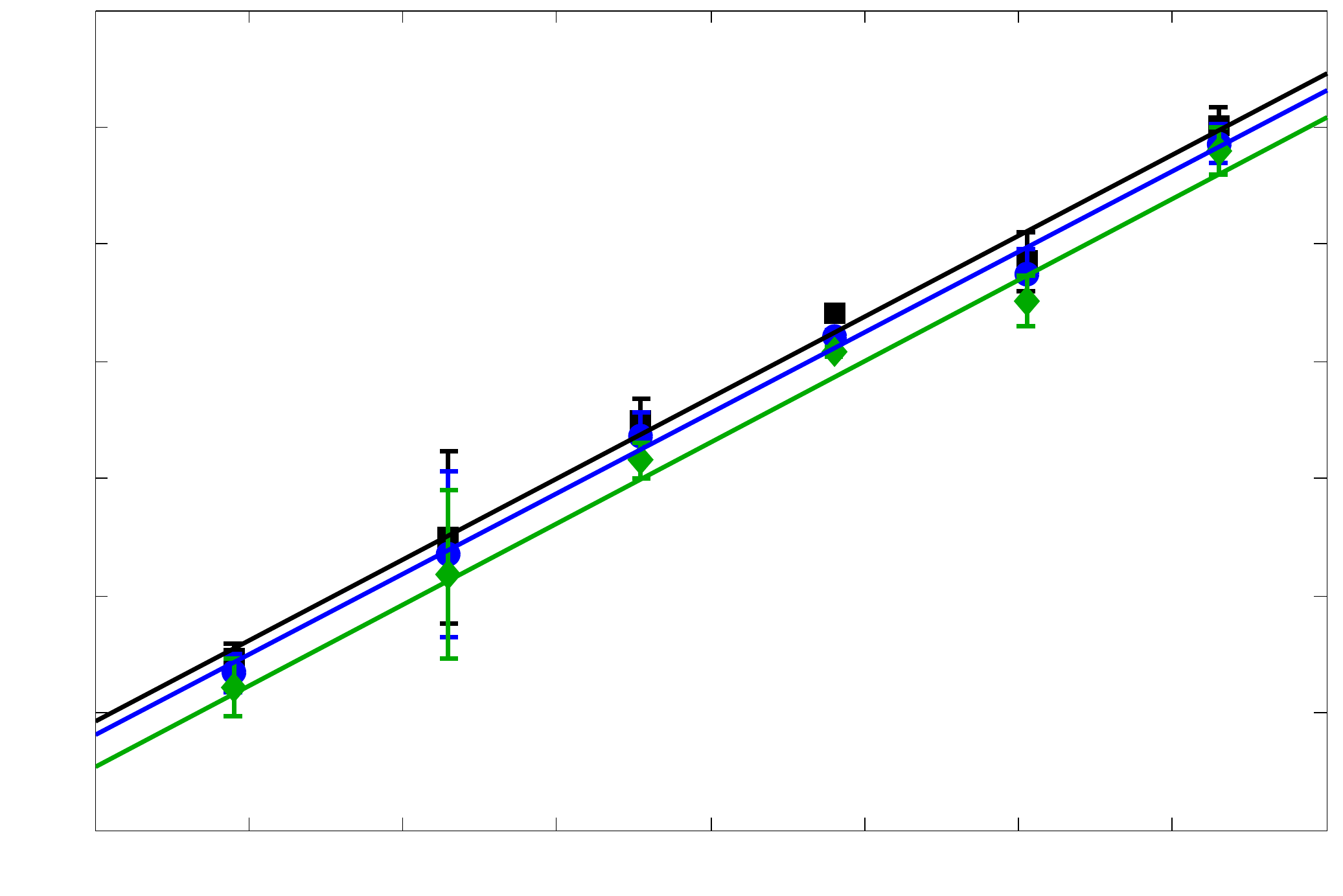}}%
    \put(0.16551003,0.002685426){\color[rgb]{0,0,0}\makebox(0,0)[lb]{\smash{50}}}%
    \put(0.27987874,0.002685426){\color[rgb]{0,0,0}\makebox(0,0)[lb]{\smash{52}}}%
    \put(0.39424913,0.002685426){\color[rgb]{0,0,0}\makebox(0,0)[lb]{\smash{54}}}%
    \put(0.50987465,0.002685426){\color[rgb]{0,0,0}\makebox(0,0)[lb]{\smash{56}}}%
    \put(0.62424336,0.002685426){\color[rgb]{0,0,0}\makebox(0,0)[lb]{\smash{58}}}%
    \put(0.73861208,0.002685426){\color[rgb]{0,0,0}\makebox(0,0)[lb]{\smash{60}}}%
    \put(0.85298247,0.002685426){\color[rgb]{0,0,0}\makebox(0,0)[lb]{\smash{62}}}%
    \put(-0.008,0.11991179){\color[rgb]{0,0,0}\makebox(0,0)[lb]{\smash{1.45}}}%
    \put(-0.008,0.29460686){\color[rgb]{0,0,0}\makebox(0,0)[lb]{\smash{1.47}}}%
    \put(-0.008,0.4693036){\color[rgb]{0,0,0}\makebox(0,0)[lb]{\smash{1.49}}}%
    \put(-0.008,0.64274188){\color[rgb]{0,0,0}\makebox(0,0)[lb]{\smash{1.51}}}%
    \put(0.23900437,-0.05423188){\color[rgb]{0,0,0}\makebox(0,0)[lb]{\smash{Concentration (\% \AmTh\ by weight)}}}%
    \put(-0.035063975,0.23815968){\color[rgb]{0,0,0}\rotatebox{90}{\makebox(0,0)[lb]{\smash{Index of Refraction}}}}%
  \end{picture}%
  \vspace{10pt}
  \caption{\label{fig:NVsc} The index of refraction of \AmThSol s increases approximately linearly with concentration (\% \AmTh\ by weight) at fixed temperatures of 17$^\circ$C (black squares), 23$^\circ$C (blue circles), and 29$^\circ$C (green diamonds). Error bars indicate the sample standard deviation of the index of refraction at a given concentration and temperature.}
\end{figure}
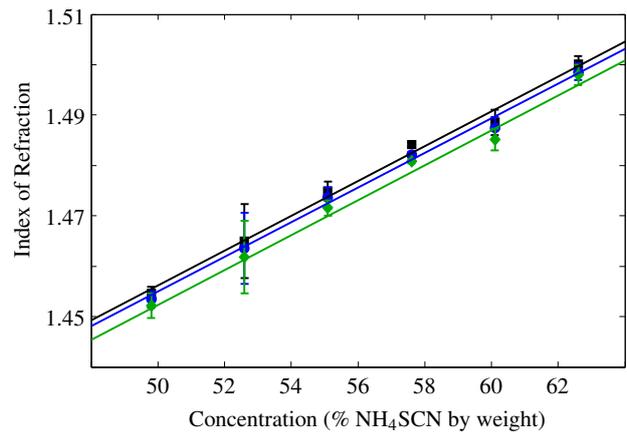

\begin{table}
\centering
  \begin{threeparttable}
    \caption{\label{tab:RefractionFitsC}Fit parameters for index of refraction versus concentration $c$ {models at} fixed temperature, $n(c) = n_2 + n_3\,c$. Here, $r^2$ is the coefficient of determination.}
     \begin{tabular}{ccccc}
      \toprule
       Temperature & $n_2$ & $n_3$ & $r^2$  \\
       ($^\circ$C)&  & (\% by weight$^{-1}$) & \\
      \midrule
					17 & 1.283 & $3.463 \times 10^{-3}$ & 0.991 \\
					23 & 1.283 & $3.440 \times 10^{-3}$ & 0.993 \\
					29 & 1.279 & $3.466 \times 10^{-3}$ & 0.991 \\      
			\bottomrule
    \end{tabular}
  \end{threeparttable}
\end{table}

Five additional data runs were taken for solutions with concentrations of 52.6\%, 57.6\%, and 60.1\% \AmTh\ by weight at temperatures of 17$^\circ$C, 23$^\circ$C, and 29$^\circ$C only. These were combined with the original data set to study the dependence of index of refraction on \AmTh\ concentration. It was found that the index of refraction increases approximately linearly with concentration at fixed temperature, as shown in Fig.~\ref{fig:NVsc}. The fit parameters for linear regression models of these data are summarized in Table \ref{tab:RefractionFitsC}.

Because the fits presented in Tables \ref{tab:RefractionFitsT} and \ref{tab:RefractionFitsC} are approximately parallel, the dependence of the index of refraction of ammonium thiocyanate solutions on temperature and concentration should be well-captured by a bivariate linear fit. Our data are well-represented by
\begin{equation}
n(c, T) = 1.2845 + 0.003513\,c - 0.0002474\,T,
\label{eq:OpticalPropertiesFit}
\end{equation} 
with 95\% of our refractive index data falling within 0.004 of the fit. Here, $T$ is the temperature in $^\circ$C and $c$ is the concentration of \AmTh\ in percent by weight. This behavior agrees qualitatively with that reported by \citet{Narrow2000} for sodium iodide solutions. 

\section{Physical properties of \AmThSol s}
\label{sec:Physical}

Unlike many other index-matching fluids, \AmThSol s have physical 
properties that are relatively close to those of water. For example,
whereas sodium iodide (\ce{NaI}) solutions used to index-match borosilicate
glass have a specific gravity of $\sim$ 1.7 and a kinematic 
viscosity of $\sim$ 2.5 cSt \citep{Narrow1998}, the equivalent \AmThSol\ has a 
specific gravity of $\sim$ 1.1 and a kinematic viscosity of $\sim$ 1.4cSt.

\begin{figure}
   \centering
  \setlength{\unitlength}{0.475\textwidth}%
  \begin{picture}(1,0.77007972)%
    \put(0,0){\includegraphics[width=\unitlength]{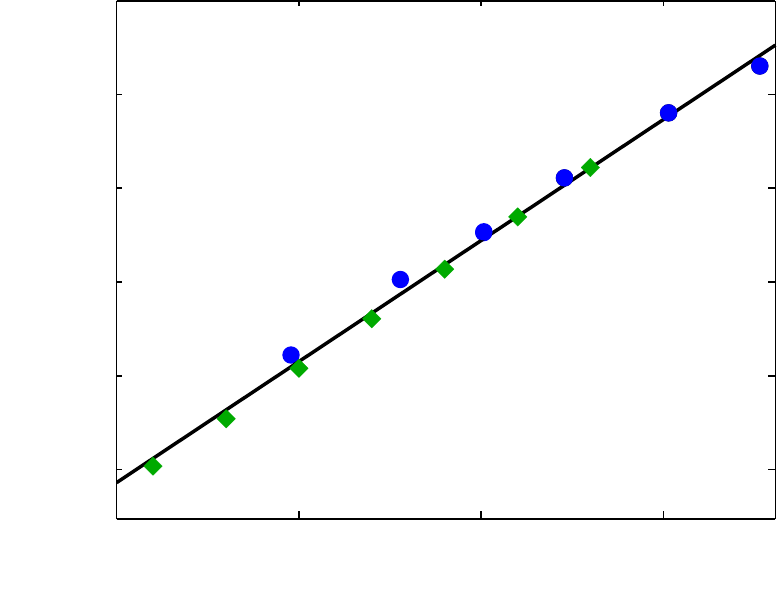}}%
    \put(0.12998087,0.05449068){\makebox(0,0)[lb]{\smash{45}}}%
    \put(0.36478946,0.05449068){\makebox(0,0)[lb]{\smash{50}}}%
    \put(0.59959358,0.05449068){\makebox(0,0)[lb]{\smash{55}}}%
    \put(0.8343977,0.05449068){\makebox(0,0)[lb]{\smash{60}}}%
    \put(0.06624832,0.15176667){\makebox(0,0)[lb]{\smash{1.10}}}%
    \put(0.06624832,0.27252308){\makebox(0,0)[lb]{\smash{1.11}}}%
    \put(0.06624832,0.39327948){\makebox(0,0)[lb]{\smash{1.12}}}%
    \put(0.06624832,0.51403589){\makebox(0,0)[lb]{\smash{1.13}}}%
    \put(0.06624832,0.63479229){\makebox(0,0)[lb]{\smash{1.14}}}%
    \put(0.27757203,0.00752985){\makebox(0,0)[lb]{\smash{Concentration (\% \AmTh\ by weight)}}}%
    \put(0.03605922,0.32619259){\rotatebox{90}{\makebox(0,0)[lb]{\smash{Density (g\,/cc)}}}}%
  \end{picture}%
  \caption{\label{fig:RhoVsC} Density as a function of \AmTh\ concentration at $\sim$23.5$^\circ$C. Our data (blue circles)
  agrees well with the historical data summarized by \citet{Washburn2003} (green diamonds). The combined data set shows an approximately linear increase in density with increasing \AmTh\ concentration, i.e., $\rho (c)=\rho_0 + \rho_1 c$.}
\end{figure}
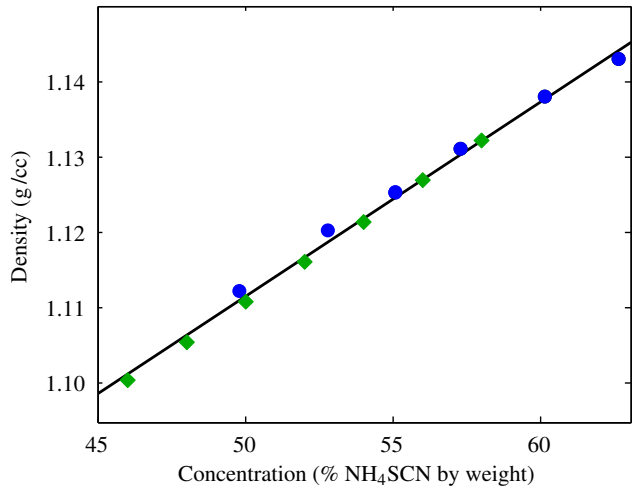

The density $\rho$ of \AmThSol s was measured as a function of concentration using an ERTCO No. 2540 hydrometer. It was determined that their density increases approximately linearly with increasing \AmTh\ concentration. Our measurements agree well with historical measurements at lower concentrations compiled by \citet{Washburn2003}. Figure \ref{fig:RhoVsC} shows both data sets along with a linear fit to the combined data, such that 
\begin{equation}
\rho (c)=\rho_0 + \rho_1 c,
\end{equation}
with $\rho_0 = 0.9824$ g/cc and $\rho_1 = 0.002583$ g/cc/\%. This fit falls within 0.002 g/cc of all the data with a coefficient of determination of $r^2 = 0.996$.

\begin{figure}
   \centering
   \setlength{\unitlength}{0.475\textwidth}
  \begin{picture}(1,0.7772286)%
    \put(0,0){\includegraphics[width=\unitlength]{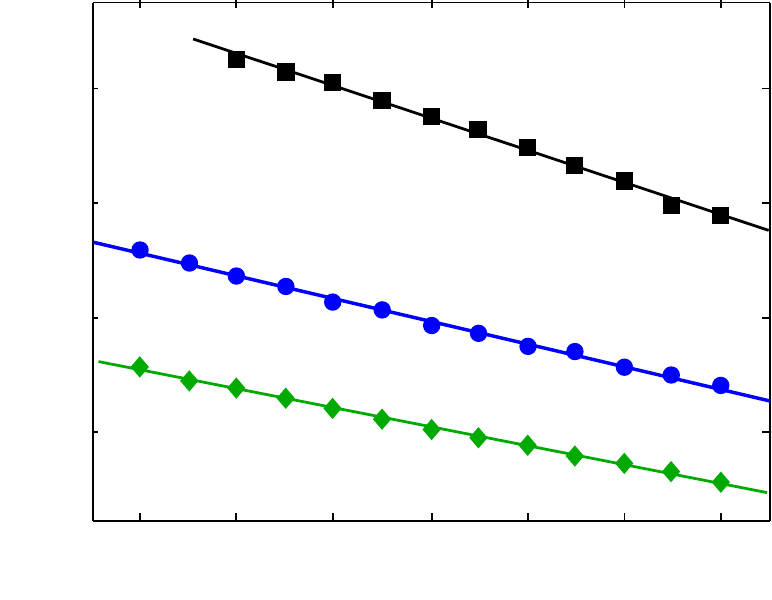}}%
    \put(0.16124166,0.05483729){\makebox(0,0)[lb]{\smash{18}}}%
    \put(0.28614187,0.05483729){\makebox(0,0)[lb]{\smash{20}}}%
    \put(0.41104658,0.05483729){\makebox(0,0)[lb]{\smash{22}}}%
    \put(0.53932247,0.05483729){\makebox(0,0)[lb]{\smash{24}}}%
    \put(0.66422268,0.05483729){\makebox(0,0)[lb]{\smash{26}}}%
    \put(0.7891229,0.05483729){\makebox(0,0)[lb]{\smash{28}}}%
    \put(0.91402311,0.05483729){\makebox(0,0)[lb]{\smash{30}}}%
    \put(0.05659554,0.20336727){\makebox(0,0)[lb]{\smash{1.2}}}%
    \put(0.05659554,0.35189725){\makebox(0,0)[lb]{\smash{1.4}}}%
    \put(0.05659554,0.50042723){\makebox(0,0)[lb]{\smash{1.6}}}%
    \put(0.05659554,0.64895721){\makebox(0,0)[lb]{\smash{1.8}}}%
    \put(0.41442226,0.00757775){\makebox(0,0)[lb]{\smash{Temperature ($^\circ$C)}}}%
    \put(0.02621441,0.24049977){\rotatebox{90}{\makebox(0,0)[lb]{\smash{Kinematic Viscosity (cSt)}}}}%
  \end{picture}%
  \caption{\label{fig:NuVsT} The kinematic viscosity of \AmThSol s with 49.8\% (green diamonds), 55.1\% (blue circles), and 62.6\% (black squares) \AmTh\ by weight decreases linearly (i.e., $\nu(T) = \nu_3 + \nu_4\,T$) with temperature at fixed concentration.}
\end{figure}
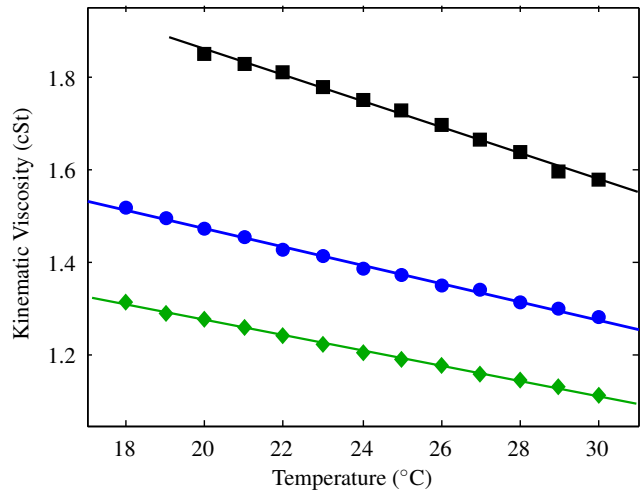

The kinematic viscosities $\nu$ of \AmThSol s having concentrations of 49.8\%, 55.1\%, and 62.6\% 
\AmTh\ by weight were measured using a No. 75 Cannon-Fenske Routine viscometer. 
The temperature was 
controlled to within $\pm$0.01$^\circ$C by immersing the viscometer in a temperature-controlled bath. As shown in Fig. \ref{fig:NuVsT},
the temperature dependence of the viscosity of these solutions is well-captured by linear fits for the range of temperatures studied. Table \ref{tab:NuFits} summarizes the fit parameters for all the solutions tested.

\begin{table}
\centering
  \begin{threeparttable}
    \caption{\label{tab:NuFits}Fit parameters for linear viscosity versus temperature {models at} fixed concentration, $\nu(T) = \nu_3 + \nu_4\,T$. $r^2$ is the coefficient of determination.}
     \begin{tabular}{ccccc}
      \toprule
       Concentration & $\nu_3$ & $\nu_4$ & $r^2$  \\
       (\% by weight)& (cSt) & (cSt\,/$^\circ$C) & & \\
      \midrule
					$62.6$ & $2.417$ & $-0.02795$ & $0.996$ \\
					$55.1$ & $1.867$ & $-0.01967$ & $0.996$ \\
					$49.8$ & $1.604$ & $-0.01633$ & $0.997$ \\      
			\bottomrule
    \end{tabular}
  \end{threeparttable}
\end{table}

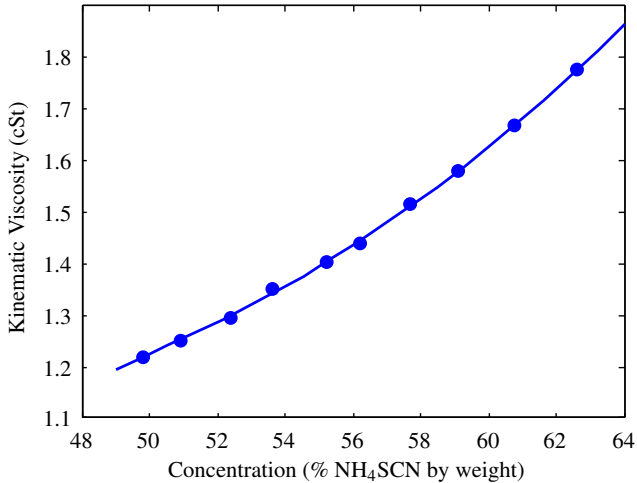
\begin{figure}
   \centering
   \setlength{\unitlength}{0.475\textwidth}
  \begin{picture}(1,0.7657449)%
    \put(0,0){\includegraphics[width=\unitlength]{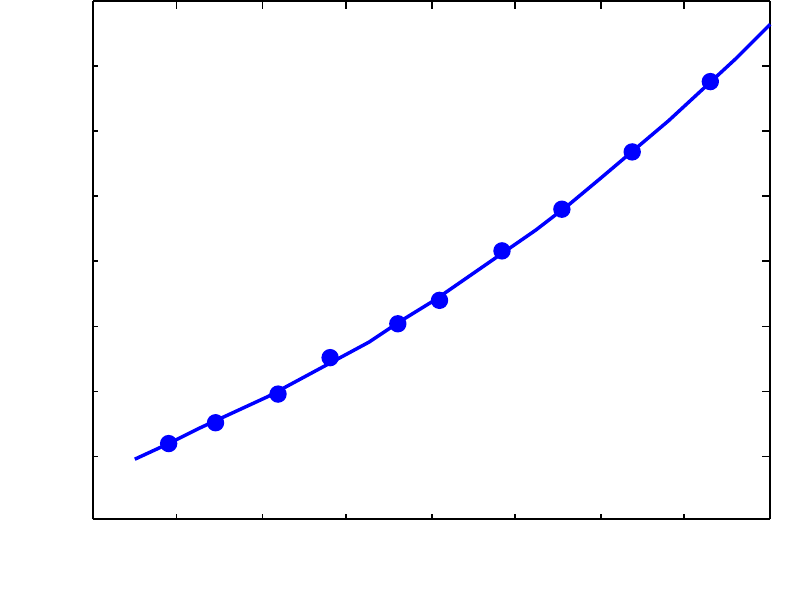}}%
    \put(0.09885168,0.05394896){\makebox(0,0)[lb]{\smash{48}}}%
    \put(0.20512359,0.05394896){\makebox(0,0)[lb]{\smash{50}}}%
    \put(0.31471649,0.05394896){\makebox(0,0)[lb]{\smash{52}}}%
    \put(0.42099283,0.05394896){\makebox(0,0)[lb]{\smash{54}}}%
    \put(0.53058573,0.05394896){\makebox(0,0)[lb]{\smash{56}}}%
    \put(0.63685764,0.05394896){\makebox(0,0)[lb]{\smash{58}}}%
    \put(0.74645054,0.05394896){\makebox(0,0)[lb]{\smash{60}}}%
    \put(0.85272245,0.05394896){\makebox(0,0)[lb]{\smash{62}}}%
    \put(0.96231535,0.05394896){\makebox(0,0)[lb]{\smash{64}}}%
    \put(0.05567872,0.09047992){\makebox(0,0)[lb]{\smash{1.1}}}%
    \put(0.05567872,0.17018385){\makebox(0,0)[lb]{\smash{1.2}}}%
    \put(0.05567872,0.25320878){\makebox(0,0)[lb]{\smash{1.3}}}%
    \put(0.05567872,0.33623371){\makebox(0,0)[lb]{\smash{1.4}}}%
    \put(0.05567872,0.41925863){\makebox(0,0)[lb]{\smash{1.5}}}%
    \put(0.05567872,0.50228356){\makebox(0,0)[lb]{\smash{1.6}}}%
    \put(0.05567872,0.58530849){\makebox(0,0)[lb]{\smash{1.7}}}%
    \put(0.05567872,0.66833341){\makebox(0,0)[lb]{\smash{1.8}}}%
    \put(0.25493855,0.007455){\makebox(0,0)[lb]{\smash{Concentration (\% \AmTh\ by weight)}}}%
    \put(0.02578975,0.23660379){\rotatebox{90}{\makebox(0,0)[lb]{\smash{Kinematic Viscosity (cSt)}}}}%
  \end{picture}%
  \caption{\label{fig:NuVsC} Kinematic viscosity as a function of \AmTh\ concentration at 23$^\circ$C (blue circles). The data is well represented by a modified Jones-Dole model of the form $\nu(c) = \nu_0 + \nu_{\nicefrac{1}{2}} \sqrt{c} + \nu_1 c + \nu_2 c^2$.}
\end{figure}

We also measured the dependence of kinematic viscosity on \AmTh\ concentration at fixed temperature (23$^\circ$C). Because the density of \AmThSol s depends only weakly on concentration in the range of concentrations studied, their kinematic viscosity is well-captured by a modified Jones-Dole model of the form 
\begin{equation}
\nu(c) = \nu_0 + \nu_{\nicefrac{1}{2}} \sqrt{c} + \nu_1 c + \nu_2 c^2,
\end{equation}
where $\nu_0 = -19.98$ cSt, $\nu_{\nicefrac{1}{2}} =$ 8.20 cSt\,/\,$\sqrt{\%}$, $\nu_1 =$ $-0.92$ cSt\,/\,\%, and $\nu_2 =$ 0.0037 cSt\,/\,\%$^2$, similar to that developed by Kaminsky for the dynamic viscosities of concentrated electrolyte solutions \citep{Kaminsky1957}. The empirical fit presented in Fig. \ref{fig:NuVsC} represents the data to better than 0.01 cSt and has a coefficient of determination $r^2$ of 0.999.

\section{Chemical Properties of \AmThSol s}
\label{sec:Chemical}

The chemical compatibility of several materials with \AmThSol s is tabulated in \citet{Washburn2003}. Standards governing these compatibility tests are a useful guide, but are designed for in\-dus\-trial applications, and we believe them to be too stringent for research applications (e.g., ASTM standards consider cosmetic changes to a material as grounds to recommend against using it). Therefore, we conducted a series of tests where samples of materials commonly used in apparatus where submerged in 55\% by weight \AmThSol\ over 19 days and observed for deterioration. 

Our tests showed that \AmThSol s are compatible with 6061 aluminum, anodized aluminum, and 316 stainless steel, as well as commonly used plastics and glass, but quickly corrode plain steel and 18-8 stainless steel. Brass and bronze become blackened but the reaction seems to stop there and the mechanical properties and finish of the parts are not significantly affected. The results of these corrosion tests are corroborated by PIV experiments in Taylor-Couette flow \citep{Borrero2014Thesis} in which \AmThSol\ was used over several years and no significant deterioration of bronze and 316 stainless steel parts was observed.

%Another important point to consider is that in order to match the index of refraction of materials like acrylic, it is necessary to use \AmThSol s that are close to saturation. In order to avoid \AmTh\ crystallizing out of the solution, it is important to stay below this limit, potentially by controlling the temperature at which the experiment is conducted. The solubility $S$ of \AmTh\ as a function of temperature can be approximated by the following third order polynomial fit to the historical data compiled by H\'{a}la \citet{Hala2004}.
%
%\begin{multline}
%S = 3.881 \times 10^{-5}\,T^3 - 2.353 \times 10^{-3}\,T^2 \\ + 0.431\,T + 54.348,
%\label{eq:solubility}
%\end{multline}
%
%\noindent
%where $T$ is temperature in $^\circ$C. This fit falls within 0.2\% by weight of all the data points and is valid for temperatures between -20$^\circ$C and 40$^\circ$C. 

Working with and handling \AmTh\ is similar to working with other refractive index-matching fluids based on heavy ionic salts like sodium iodide. Handling of powdered ammonium thiocyanate should be done with care and appropriate protective equipment (gloves, dust mask, safety goggles), as it is toxic if inhaled or ingested (GHS category 4) and can be a skin irritant. Once in solution, however, it is easy to work with safely as long as it is contained within the experimental apparatus. Appropriate care should be taken when disposing \AmTh\ solutions, as they could be potentially harmful to aquatic life (GHS Category 3).  

\section{Summary}
\label{sec:Summary}
We have shown that \AmTh\ solutions provide an alternative to common refractive index-matching fluids and presented an empirical model for $n$ as a function of temperature and \AmTh\ concentration. Our data show that these solutions can be used to match the index of refraction of fused quartz, borosilicate glass, and acrylic. They are also chemically compatible with other materials frequently used to build experimental apparatus. Most uniquely, their viscosities and densities are close to those of water, which make them useful working fluids for researchers carrying out optical measurements at high Reynolds numbers.

\begin{acknowledgements}
The authors would like to thank M.F. Schatz, who supported early stages of this work with funds from National Science Foundation grant \# CBET-0853691. Additional support for this project was provided by the Summer Scholarship Fund and the James Borders Physics Student Fellowship at Reed College. The authors also thank M. Yoda, who provided the hydrometers used in this work. \end{acknowledgements}

\bibliographystyle{spbasic}      % basic style, author-year citations

\end{document}